# A Database Engineered System for Big Data Analytics on Tornado Climatology

Fengfan Bian[1], Carson K. Leung[1] ✉ [0000-0002-7541-9127], Piers Grenier[1],
Harry Pu[1], Samuel Ning[1], and Alfredo Cuzzocrea[2,3]

[1] University of Manitoba, Winnipeg, MB, Canada
[2] University of Calabria, Rende, Italy
[3] Université Paris Cité, Paris, France
✉ Carson.Leung@UManitoba.ca

**Abstract.** Recognizing the challenges with current tornado warning systems, we investigate alternative approaches. In particular, we present a database engineered system that integrates information from heterogeneous rich data sources, including climatology data for tornadoes and data just before a tornado warning. The system aids in predicting tornado occurrences by identifying the data points that form the basis of a tornado warning. Evaluation on US data highlights the advantages of using a classification forecasting recurrent neural network (RNN) model. The results highlight the effectiveness of our database engineered system for big data analytics on tornado climatology—especially, in accurately predicting tornado lead-time, magnitude, and location, contributing to the development of sustainable cities.

**Keywords:** Database Engineered Application, Data Science, Data Mining, Data Analytics, Climate Data, Tornado.

## 1   Introduction

In the current era of big data, data science or database engineered systems [1, 2]—which make good use of machine learning (ML) [3, 4] (including deep learning [5, 6]), data mining [6-14] and analytics [15-17]—help reuse and/or integrate information in various real-life applications for public good. These include healthcare informatics [18-22] and social network analysis [23, 24].

In this paper, we focus on climate data analysis—in particular, tornado climatology. The tornado is a highly destructive, life-threatening natural disaster that can cause extensive damage to buildings, infrastructure, and agriculture resulting in significant property loss, economic hardship, as well as tragically, the loss of life. Tornadoes are most common in the USA than in any other countries. On average, there are about 1,200 tornadoes (including confirmed and unconfirmed) that touch down on the USA each year. Not only the USA has the highest quantity of tornado, it also has the highest quantity of intensive tornado. For instance, violent tornadoes—e.g., those rated EF4 or EF5 on the Enhanced Fujita scale (i.e., wind speed ≥ 166 mph) leading devastating or incredible damage—occur most often in the USA than in any other countries.

Current tornado warning systems provide only a limited lead-time for residents and authorities to prepare an effective response. Improving the accuracy and lead-time of tornado prediction is crucial for saving lives as well as minimizing damage. In the USA, tornadoes occur quite frequently and clearly, a specialized forecasting system is needed. These tornadoes often occur in regions with lower population density, making their impact on communities less predictable. As a result, the current warning system, while valuable, lacks the granularity required for precise preparedness and response.

Machine learning is commonly employed to create this desired forecasting system, where the model typically incorporates a neural network. Here, the pre-storm environment is used to train the network and prediction involves classifying weather data points based on if they constitute a tornado event. This provides the motivation for this paper as it present a database engineered system—namely, a recurrent neural network (RNN) model for tornado forecasting.

When considering the benefits of a predictive forecasting system, it is important to acknowledge its potential real-life applications. Examples include:

- advancement in comprehensive information systems for tornado genesis,
- safe evacuation procedures to minimize casualties,
- standardization of prompt warning systems to increase tornado prediction lead-time,
- tornado protection for a broader range of regions, and
- real-time analyses on tornado speed and strength.

These are just some of the real-world database engineered applications of using a predictive forecasting model, and the advantages they offer are clear.

In this paper, we present a database engineered system to integrate information from multiple rich data sources such as climatology data for tornadoes and those immediately preceding a tornado warning. The resulting system helps predict the occurrence of tornadoes. It determines the data points that constitute a tornado warning. Our *key contributions* of this paper include:

- a new long short-term memory (LSTM) RNN predictive forecasting model with 85.94% cross-validation accuracy;
- active learning for labeling weather data points and dynamically updating when new data arrives, which is invaluable when considering the issuing of tornado warnings;
- classification of new data points to predict a tornado event or not;
- implementation of 10-fold cross validation for multiple training sets to reduce bias in results by testing on a more diverse range of data; and
- new insights for yearly model accuracy, the area under the curve (AUC) scores for the receiver operating characteristic (ROC) curve in each fold, a confusion matrix for evaluating our classification model, and model training accuracy vs. loss.

The remainder of this paper is organized as follows. The next section discusses some background and related works. Section 3 presents our database engineered system. Sections 4 and 5 shows our evaluation results and draws conclusions, respectively.

## 2  Background and Related Works

Existing works on tornado forecasting systems usually use the following approaches:

- statistical model [25],
- active learning with support vector machines [26],
- machine learning classifiers [27],
- deep learning [28-31],
- fuzzy logic and classification [32],
- feature extraction classification [33], or
- spatiotemporal relational probability trees [34].

However, the statistical model can only make predictions in a short-term diagnostic mode because it does not incorporate active learning techniques, and instead uses passive learning [25]. Consequently, less significant data points for distinguishing a tornado event are labeled and placed into the training set for their model. In contrast, our database engineered system is naturally different because it does not incorporate passive learning, where predictions are made using old data points. In other words, the statistical model is static in that it labels data points randomly during training and it may not be able to update dynamically with new data.

The classifiers for the imbalanced tornado data model deal with the minority class, in which tornadoes form from unusual pre-storm conditions as represented by the weather data [26]. The objective being to maximize the accuracy of predicting tornadoes in the rare cases, which naturally makes the classifiers less applicable for standard tornado occurrences. In contrast, our database engineered system deals with the majority class for tornado prediction.

Most works involving deep learning for predicting tornado events use convolutional neural networks (CNNs) [28-30], which are multi-layered and designed to learn from spatial grids. These spatial grids must first be converted into scalar features, which can then be used to train the network. The model can then predict the outcome of unseen data using training knowledge. Due to the complexity of the data, CNNs require a very large number of samples to learn from. They also need heavy tuning as well as hardening to prevent false-positives, and especially false-negatives. McGuire and Moore [30] tested a variety of CNNs based "on their ability to classify the strength of daily tornado outbreaks". The relevant CNNs include LeNet-5, VGG-16 and Resnet-50, which all use object recognition in image datasets. Soni et al. [31] implemented a counterfeit neural network (ANN) where weather parameters can access future information on weather history. This means that the parameters of temperature, rain and air speed are expected without a lot of flexibility. In contrast, our database engineered system uses an RNN. The main difference being that an RNN has fewer layers and requires much less data for learning. In addition, our system does not compare multiple neural networks for tornado prediction, and is therefore different from related works that predict tornado days in the USA. Moreover, we do not allow parameter access to future information on weather history, and is therefore more flexible at the cost of accuracy.

Lakshmanan et al. [32] incorporated feedforward neural networks and support vector machines to automatically classify radar signatures for tornado prediction. The classification is accomplished through an integration of gradient estimation, fuzzy logic into the classifier. As such, a more probabilistic approach was taken as opposed to a machine learning one by viewing tornado prediction as a spatiotemporal problem "of estimating the probability of a tornado event at a particular spatial location within a given time window". Consequently, smaller timeframes was resulted for predictions on tornado events. In contrast, our database engineered system uses an RNN.

Coccomini and Zara [33] collected meteorological data from certain geological locations daily and placed them into climate metrics. Features were then extracted and stored as single-day features and multiple day features for classification. Their model had 84% accuracy for up to 5 days of prediction, which could be more accurate. Although our database engineered system is similar in the sense that it also aims to develop accurate early-detection tornado systems, it is different in its temporal analysis of the collected data and the features extracted

McGovern et al. [34] used probability estimation trees, which learn using data temporally varies (also known as spatiotemporal relational probability trees, or SRPTs for short). Their probabilistic model uses SRPTs to automatically discover spatial temporal relationships in data. They also only discusses how to evaluate a machine learning model rather than incorporating one, and only covers prediction of tornadoes in state of Oklahoma, USA. In contrast, our database engineered system focuses on the USA as a whole and uses a machine learning approach.

## 3 Our Database Engineered System

### 3.1 Overview

Our database engineered system aims to address these challenges by designing and developing a tornado forecasting system that accounts for geographical and meteorological characteristics. By utilizing tornado data, meteorological parameters and advanced machine learning techniques, we aim to create a localized forecasting system capable of providing detailed forecasts—including specific timing, magnitude and geographical location of tornado events.

Our system also uses data mining to make tornado forecasting better. We use machine learning to look at old weather data and find patterns that might tell us when and where tornadoes will happen. Our method differs from related works as we implement a classification predictive model. First, we group the data based on the kind of weather that often precedes tornadoes and then use classification to figure out which of these weather groups usually lead to tornadoes. This helps users know information like how strong the tornado will be, where it will happen, and when it might hit. Using these data mining techniques enables residents to have more time to get ready for tornadoes, which can help keep them safe.

We focus on utilizing existing datasets sourced from the USA, where most tornadoes occur east of the Rocky Mountains, the Great Plains, the Midwest, the Mississippi Valley and the southern USA. Among the 48 continental US states, although Florida has the highest number of tornadoes per unit area, these are usually weak tornadoes. In contrast, Oklahoma has the highest number of *strong* tornadoes. With warm moist air from southeast (Gulf of Mexico) and warm dry air from southwest (Gulf of California and Equator in the Pacific Ocean) meeting cold dry air from the northwest (and Canadian Rockies), Tornado Alley—covering parts of Texas, Oklahoma, Kansas, Colorado, Nebraska and South Dakota—becomes an ideal environment for tornadoes to form within developed thunderstorms.

In our presented tornado forecasting system, we employ a binary classification approach integrated into an RNN architecture to overcome forecasting challenges. Unlike traditional feedforward neural networks that process input data in a single pass, RNNs handle sequential data by maintaining a hidden state that captures information about previous inputs. This sequential memory makes RNNs particularly suitable for time-series forecasting, such as tornado prediction, where the temporal order of meteorological events is crucial. The system leverages the capabilities of the RNN to discern patterns indicative of tornado occurrence based on meteorological data.

In the context of tornado classification prediction, RNNs offer advantages over other models due to their ability to capture temporal dependencies and patterns in the data. Our presented database engineered system for tornado forecast utilizes a binary classification approach integrated into an RNN architecture, allowing it to overcome forecasting challenges effectively. Tornado events unfold over time, with meteorological conditions evolving dynamically, and RNNs excel at capturing the temporal nuances in such sequences. By considering the sequential nature of meteorological parameters leading up to a tornado event, RNNs can potentially capture patterns that other models (e.g., those treating the data as independent samples) may fail to recognize. In addition, RNNs are well suited for handling varying lengths of input sequences, a critical feature when dealing with meteorological data with irregular time intervals or different sampling rates. This flexibility allows the model to adapt to the diverse temporal aspects of tornado formation, enhancing its ability to generalize across different scenarios. While other machine learning models—e.g., traditional feedforward neural networks, support vector machines (SVM)—may perform well in certain applications, their limitations in handling sequential data make them less suitable for tornado prediction tasks. In contrast, the RNN's capacity to learn and remember temporal dependencies positions it as a promising choice for capturing the complex dynamics associated with tornado formation.

### 3.2 Design Details

We integrate information about USA weather and tornado. Within these data, we explore the following key features:

- Temperature metrics: maximum, minimum, and average temperatures.
- Perceived temperature.

- Moisture and dew: dew point, and the amount of moisture in the air.
- Precipitation: amount of precipitation, its probability, and its coverage.
- Wind information: wind speed and direction.
- Pressure and atmospheric conditions: atmospheric pressure at sea level.
- Cloud and visibility: cloud coverage, and the distance of visibility.
- Moon phase: phase of the moon.

As for the output of the model, it is designed for binary classification—specifically to predict whether a tornado will occur or not. The binary output variable is denoted as result, where a value of 1 indicates the occurrence of a tornado, and 0 signifies the absence of a tornado.

The RNN training process involves splitting the data into features and labels. We then standardize the numerical features (say, by using the Scikit-learn's StandardScaler) and reshape the data to suit the RNN format. The Synthetic Minority Oversampling Technique (SMOTE) is applied to address potential imbalances in the target variable. Moving on to the model architecture, the RNN model incorporates:

- one LSTM layer,
- a dropout layer for regularization, and
- a dense output layer with a sigmoid activation function.

Then, the model utilizes a 10-fold cross-validation approach for training and evaluation. This involves dividing the data into 10 subsets, training the model on 9 of them, and validating the remaining subset. This process repeats 10 times, and accuracy scores for each fold are collected. To prevent overfitting, a dropout rate of 0.2 is applied during the model creation. Lastly, the overall model performance is assessed by computing the mean and standard deviation of the accuracy scores across all folds.

### 3.3  Implementation Details

To prepare and clean data, we utilize the Python libraries Pandas (for data analysis) and NumPy (for supporting large multi-dimensional arrays). We meticulously process the weather data, discarding non-numeric and sparsely populated columns that could potentially skew our analyses. The "datetime" conversion is not just a procedural necessity but also a strategic move to ensure the sequential integrity of our dataset, which is vital for time-series analysis. By imputing the missing values with the mean for the numeric columns, we preserve every data point's potential to contribute to our findings.

Selecting the RNN as our predictive model is a deliberate choice, driven by the sequential and temporal nature of meteorological data. The RNN's innate ability to process sequences through its internal state and memory capabilities makes it an unparalleled tool for identifying the precursors to tornado events. The preprocessing phase involves not just standardizing the data but also transforming it into a format amenable to RNN processing. Addressing data imbalance with the SMOTE is also critical, as it ensures our model not being biased towards the majority class and can generalize well across all scenarios.

Our model's architecture is thoughtfully designed using Keras (an open-source library that provides a Python interface for ANNs) and incorporating LSTM layers (which are renowned for their sequence prediction capabilities). We introduce dropout layers as a precautionary measure against overfitting, and the binary nature of the output layer directly corresponded to the occurrence or non-occurrence of tornado events. The robustness of our model is further bolstered by a comprehensive 10-fold cross-validation, which ensures that every data point is utilized in both training and validation phases, thus minimizing any potential bias and enhancing the model's applicability across diverse data samples.

The coding implementation was a testament to the end-to-end process, from initial data cleaning to the final stages of model evaluation. By deploying Scikit-learn's KFold cross-validator, we not only divide our dataset for validation purposes but also set up a robust framework for tracking and analyzing the model's accuracy and AUC across different folds. This attention to detail allows our model's predictive performance to be fine-tuned and ensure its reliability. The resultant confusion matrices provide a cumulative view of the model's true positive and negative rates, presenting a clear picture of its predictive prowess. Moreover, the visual aids (e.g., ROC curves, plots) juxtaposing training accuracy with loss, served as intuitive indicators of the model's learning trajectory and performance benchmarks.

## 4 Evaluation

### 4.1 Setup

Our investigation into tornado prediction leverages a significant corpus of meteorological data amassed from Visual Crossing Weather Data Services[1]. This platform is a repository of historical weather data, meticulously curated to aid a wide array of climatological research. The service provides a wealth of information essential to the analysis—covering various weather-related parameters such as temperature, precipitation, and wind, which are critical factors in understanding and forecasting the conditions that lead to tornado occurrences. The granularity and precision of the data from Visual Crossing enabled our model to capture the nuanced environmental patterns that typically precede tornado events.

In conjunction with the Visual Crossing dataset, we incorporated the Storm Prediction Center (SPC) dataset[2] provided by the National Oceanic and Atmospheric Administration (NOAA). Renowned for its comprehensive documentation of severe weather events, the SPC dataset is particularly detailed in its accounts of tornado occurrences, including geographical and temporal markers. These data are invaluable for training a predictive model, as they allow for the calibration of outputs against well-documented historical instances of tornadoes, thus enhancing the model's predictive precision with respect to location and intensity.

---

[1] https://www.visualcrossing.com/
[2] https://www.spc.noaa.gov/

The datasets collectively provide a robust analytical foundation, comprising over 178,997 records post-cleanup, each encompassing around 28 attributes reflecting various meteorological conditions. See Table 1 for yearly distribution of data and number of tornado occurrences.

Table 1. Yearly distribution of data and number of tornado occurrences

| Year | Number of rows | Number of tornado |
|---|---|---|
| 1998 | 17,885 | 491 |
| 1999 | 17,885 | 425 |
| 2000 | 17,934 | 413 |
| 2001 | 17,885 | 405 |
| 2002 | 17,885 | 336 |
| 2003 | 17,885 | 398 |
| 2004 | 17,934 | 489 |
| 2005 | 17,885 | 380 |
| 2006 | 17,885 | 363 |
| 2007 | 17,885 | 381 |

Our system uses a RNN model, meticulously designed to process and learn from the intricate sequences inherent in weather data. The model was designed to classify data points for predicting the likelihood of tornado events. It uses a binary "result" output where a value of "1" signifies the prediction of a tornado. Through the use of the RNN, the study harnessed the power of sequential data processing to uncover patterns across the temporal landscape of weather conditions that culminate in tornado genesis.

The datasets from Visual Crossing and the SPC are indispensable in the development of our system. The diverse and high-dimensional dataset enables the construction of a predictive model that not only aims to enhance the accuracy of tornado forecasts but also potentially extends the lead-time for warnings, thereby contributing to disaster preparedness and saving lives. The quality and depth of the datasets reflect the current state-of-the-art in meteorological data collection and severe weather forecasting, underscoring the value of reliable data sources in advancing the field of climatological research.

To evaluate and demonstrate the usefulness of our database engineered system, we test it with 10-fold evaluation. We randomize the data before splitting it into 10 folds, which is good for ensuring that each fold is representative of the whole. We set the random state of the 10-fold evaluation to 42 to ensure the reproducibility of our results. From the 10-fold validation results, we analyze the results using AUC score, confusion matrix and model training accuracy and loss.

### 4.2 Evaluation Results

Figure 1 shows the AUC scores for our ROC curve from the 10-fold cross-validation process. The AUC scores range from just under 0.856 to just over 0.864. This indicates a fairly consistent model performance across different folds as the variation is within a narrow band. There is a noticeable dip in AUC score in fold 6, which suggests that the data in fold 6 was different from the rest of the folds, possibly due to outliers in fold 6.

Despite the variability, The AUC scores are quite high as they are all above 0.85, suggesting our model has a good ability to differentiate between the positive class (presence of tornado) and negative class (absence of tornado). Apart from the dip in fold 6, the model's performance is relatively consistent, which proves the robustness of our model. Overall, the AUC scores suggest our model has good discrimination to distinguish positive and negative outcomes of tornados.

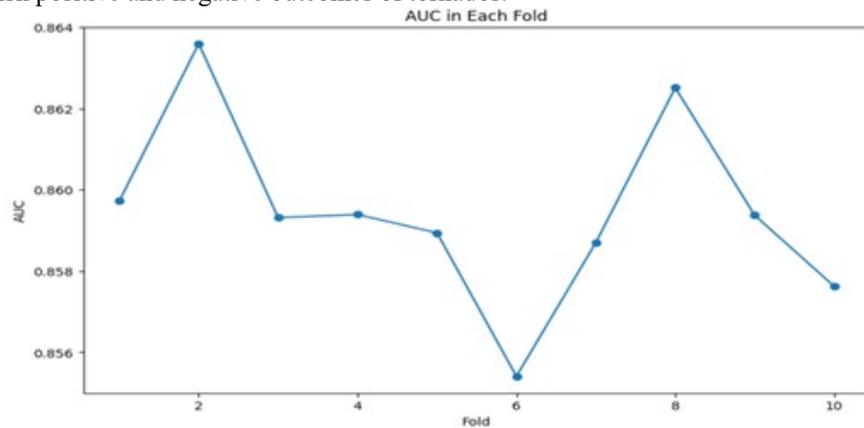

**Fig. 1.** AUC in each fold for 10-fold cross validation.

Figure 2 shows the confusion matrix table of our 10-fold cross validation results. The row of the matrix represents the instances of an actual class from the results, while the column represents the instances in a predicted class. Here are the results analysis: Our model correctly predicted the negative class approximately 14,412 times on average. The model also correctly predicted the positive class approximately 15,644 times on average.

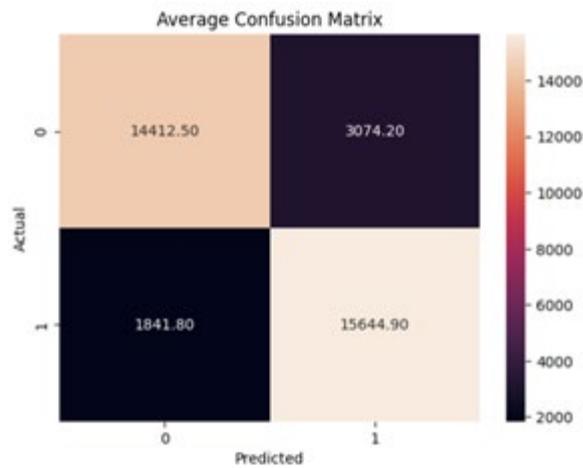

**Fig. 2.** Average confusion matrix for 10-fold cross validation results.

Based on the values in this confusion matrix, we compute the metrics shown in Table 2:

$$\text{Sensitivity} = \text{true positive rate (TPR)} = \frac{TP}{TP+FN} \quad (1)$$

$$\text{False positive rate (FPR)} = \frac{FP}{FP+TN} \quad (2)$$

$$\text{Accuracy} = \frac{TP+TN}{TP+TN+FP+FN} \quad (3)$$

$$\text{Precision} = \text{positive predictive value (PPV)} = \frac{TP+TN}{TP+TN+FP+FN} \quad (4)$$

$$\text{Negative predictive value (NPV)} = \frac{TN}{TN+FN} \quad (5)$$

where TP is true positive, TN is true negative, FP is false positive, and FN is false negative. For instance:

- High sensitivity (0.895): The model demonstrates strong sensitivity, correctly identifying approximately 89.5% of the positive cases. This indicates a high true positive rate, suggesting that the model is effective in detecting the presence of the condition or characteristic being tested.
- Moderate false positive rate (FPR) (0.176): With a FPR of 17.6%, there is a noticeable, but not overwhelming, tendency for the model to incorrectly classify negative cases as positive. This level suggests room for improvement in distinguishing non-cases more accurately.
- Good accuracy (0.859): An accuracy of 85.9% signifies that the model correctly classifies both positive and negative cases with high reliability in the majority of instances. This reflects its overall effectiveness in making correct predictions.
- Solid precision (0.836): The precision of 83.6% indicates that when the model predicts a positive result, there is an approximately 83.6% chance that it is correct. This points to the model's reliability in its positive predictions, though there is still some scope for reducing false positives.
- High negative predictive value (NPV) (0.887): An NPV of 88.7% shows that the model is quite effective in predicting negative cases. When the model predicts a negative result, it is correct about 88.7% of the time, which is a strong indication of its ability to correctly identify true negatives.

Table 2. Metrics based on the confusion matrix

| Metric | Equation | Value |
|---|---|---|
| Sensitivity aka true positive rate (TPR) | TP / (TP+FN) | 89.5% |
| False positive rate | FP / (FP+TN) | 17.6% |
| Accuracy | (TP+TN) / (TP+TN+FP+FN) | 85.9% |
| Precision, aka positive predictive value (PPV) | TP / (TP+FP) | 83.6% |
| Negative predictive value (NPV) | TN / (TN+FN) | 88.7% |

In summary, the confusion matrix and the derived metrics suggest that the model is quite effective in identifying true positives and true negatives when it comes to predict-ing tornadoes, with a high degree of sensitivity and accuracy. The moderate false positive rate and good precision indicate that while the model is reliable in its predictions, there is potential for further refinement, particularly in reducing the number of false positives. Overall, the model demonstrates a robust performance in classification tasks, making it a valuable tool for tornado prediction.

Figure 3 shows the model training accuracy and loss. The graph illustrates the model's performance during the training phase, measured by accuracy and loss over a series of 10 epochs. Notably, the accuracy curve (in blue) ascends sharply at the onset and begins to level off as it approaches the final epochs, indicating that the model is making increasingly correct predictions as it processes more training data. The plateau towards the end of the training suggests that the model may be nearing its peak learning capacity given the current data and network architecture.

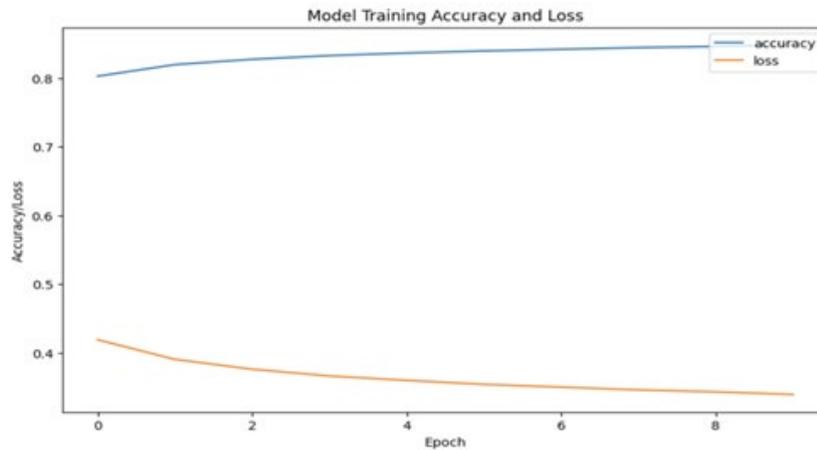

**Fig. 3.** Model training accuracy and loss.

Conversely, the loss curve (in orange) demonstrates a significant decrease, particularly evident in the initial epochs, reflecting a reduction in the error between the model's predictions and the actual targets. The loss curve's decline is steeper in the early stages of training, which is common as the model quickly corrects itself from initial random or uninformed predictions. As the epochs progress, the rate of loss reduction slows, suggesting that the model is starting to converge and is making fewer and smaller mistakes in its predictions.

The convergence of these two metrics—accuracy and loss—towards their respective high and low values indicates a successful training process. However, the graph also implies that further training beyond the 10th epoch may not yield substantial improvements and could lead to overfitting, where the model learns patterns specific to the training data that do not generalize well to unseen data.

# 5      Conclusions

In conclusion, this paper addresses the critical issue of tornado prediction, considering the high frequency of tornado occurrences in the USA and the limitations of existing warning systems. The motivation stems from the need to enhance the accuracy and lead-time of tornado prediction to minimize the devastating impact on lives and property. The presented database engineered system utilizes an LSTM-RNN predictive forecasting model, achieving an impressive 85.94% k-fold cross-validation accuracy. The real-life applications of such a predictive forecasting system are extensive, ranging from comprehensive information systems for tornado genesis to safe evacuation procedures, standardizing warning systems, and providing tornado protection for a broader range of regions. Our key contributions of this paper include active learning for labeling weather data points, dynamic updates with new data, classification of data points for tornado prediction, 10-fold cross-validation for diverse training sets, and insights into model accuracy metrics.

Comparisons with existing models underscore the uniqueness of the proposed LSTM RNN model in our database engineered system. For instance, our system incorporates active learning and dynamically updates with new data. It focuses on the majority class for tornado prediction, distinguishing it from models targeting imbalanced tornado data. While sharing similarities with deep learning CNN models, our RNN model offers advantages in terms of reduced complexity and data requirements. Moreover, the model prioritizes flexibility over accuracy. The existing spatiotemporal method for predicting tornado events (which views tornado prediction as a spatiotemporal problem) differs significantly from our RNN model, which employs machine learning for classification. Our system also contains a tailored tornado forecasting system for the USA, addressing challenges posed by unique geographical and meteorological characteristics. Leveraging US tornado data and advanced machine learning techniques, our system employs a binary classification approach integrated into an RNN architecture. This approach proves advantageous in capturing temporal dependencies and patterns crucial for tornado prediction, surpassing the limitations of traditional models in handling sequential meteorological data.

The model, trained on a dataset combining USA weather and tornado data, focuses on key meteorological parameters to predict tornado occurrence using a binary output variable. The methodology involves rigorous data cleaning, RNN model selection, and careful consideration of imbalances using SMOTE. The 10-fold cross-validation approach ensures robust training and evaluation, emphasizing the model's ability to generalize across diverse data samples. Datasets from Visual Crossing and the Storm Prediction Center (SPC) provide a comprehensive foundation for the model, capturing nuanced environmental patterns preceding tornado events. The model's evaluation involves AUC graph analysis, confusion matrix interpretation, and examination of model training accuracy and loss. The AUC scores consistently above 0.85 demonstrate the model's strong discrimination between positive and negative tornado outcomes. Confusion matrix analysis reveals the model's high sensitivity, moderate false positive rate, good accuracy, solid precision, and high negative predictive value, indicating effectiveness in identifying true positives and true negatives. The model's training accuracy and loss graphs suggest successful convergence, with potential caution against overfitting beyond the 10th epoch. Our database engineered system highlights a robust tornado forecasting system with promising predictive capabilities.

As *ongoing and future work*, we would integrate other relevant data with an aim to further enhance prediction accuracy and data analytics. This would further enhance our system in classifying tornado climatology data and predicting future disasters for sustainable cities. We would also incorporate OLAP techniques (e.g., [35-39]) in our system.

**Acknowledgments.** This work is partially supported by NSERC (Canada) and University of Manitoba.